\documentclass[aps,preprint,prd,showpacs,nofootinbib]{revtex4}

\usepackage{booktabs}
\usepackage{latexsym}
\usepackage{bm}
\usepackage{appendix}
\usepackage{amsmath,amssymb,tabularx}
\usepackage{graphicx,subfigure}
\usepackage{color}
\usepackage{booktabs}
\usepackage[T1]{fontenc}
\usepackage{float}
\usepackage[colorlinks,linkcolor=magenta,anchorcolor=cyan,citecolor=blue]{hyperref}

\begin{document}

\title{Bounded islands in dS$_{2}$ multiverse model }

\author{Wen-Hao Jiang$^{1}$\footnote{jiangwenhao16@mails.ucas.ac.cn}}
\author{Yun-Song Piao$^{1,2,3,4}$\footnote{yspiao@ucas.ac.cn}}

\affiliation{$^1$ School of Physical Sciences, University of
Chinese Academy of Sciences, Beijing 100049, China}

\affiliation{$^2$ School of Fundamental Physics and Mathematical
    Sciences, Hangzhou Institute for Advanced Study, UCAS, Hangzhou
    310024, China}

\affiliation{$^3$ International Center for Theoretical Physics
    Asia-Pacific, Beijing/Hangzhou, China}

\affiliation{$^4$ Institute of Theoretical Physics, Chinese
    Academy of Sciences, P.O. Box 2735, Beijing 100190, China}

\begin{abstract}

The cosmological event horizons are observer-dependent, which
might bring a paradox. As an example, in dS$_{2}$ multiverse model
there are entanglement islands in crunching regions encoding the
information of regions near future infinity of inflating or
Minkowski bubbles, however, for two observers in different
bubbles, since their island regions overlap, both observers will
be able to get access to the information encoded in the
overlapping region, indicating a violation of no-cloning theorem.
In this paper, we present a different resolution to this paradox.
Based on the Petz $\mathrm{R\acute{e}nyi}$ mutual information, we
show that besides the quantum extremal surfaces there might be
another boundary for the island in corresponding spacetime so that
the island regions are bounded by ``division points" rather than
extending to the rest of the entire spacetime. We also discuss the
implications of our result.

\end{abstract}

\maketitle

\section{Introduction}\label{1}

Black hole is an ideal object to study quantum gravity, see e.g.
recent Ref.\cite{Bousso:2022ntt}. Recently, one of the most
notable achievements is the derivation of Page Curve
\cite{Page:1993wv,Page:2013dx}, the generalized entropy of radiation
$S_R$ follows the 'island formula'
\cite{Penington:2019npb,Penington:2019kki,Almheiri:2019psf,Almheiri:2019hni,Almheiri:2019qdq,Almheiri:2019psy,Almheiri:2019yqk},
which is regarded as a significant step toward resolving the black
hole information paradox \cite{Hawking:1975vcx,Hawking:1976ra}:
\begin{equation}\label{eq1.1}
    \mathit{S}_{\mathit{R}}=\mathrm{min} \{\mathrm{ext}[\frac{A(\partial I)}{4G_N}+S_{\mathrm{semi-cl} }(\mathrm{Rad} \cup I)]\},
\end{equation}
where $I$ is the island region surrounded by quantum extremal
surface (QES) \cite{Engelhardt:2014gca}, see also
\cite{Ryu:2006bv,Hubeny:2007xt,Lewkowycz:2013nqa,Faulkner:2013ana,Wall:2012uf},
which appears after Page time, entangling with the radiation
emitted by black hole. The occurrence of island suggests that the
information of a region inside the horizon can be encoded in the
state of another system outside the horizon.
%\cite{Almheiri_2020,hollowood2020islands,Harlow_2016,almheiri2023islands,penington2020entanglement,Almheiri_2021},
As event horizons also exist in cosmology, it is interesting to
consider the implication of island for cosmology, see e.g.recent
\cite{Hartman:2020khs,Chen:2020tes,Balasubramanian:2020xqf,Aguilar-Gutierrez:2021bns,Levine:2022wos,Piao:2023vgm,Yadav:2022jib,Espindola:2022fqb,Ben-Dayan:2022nmb,Kames-King:2021etp,Aalsma:2021bit,Baek:2022ozg,Aalsma:2022swk,Teresi:2021qff,Seo:2022ezk,Azarnia:2021uch,Choudhury:2020hil,Choudhury:2022mch,Aguilar-Gutierrez:2023zoi}
\footnote{The applications of island rule have been studied
intensively in varity of spacetimes, see
e.g.Refs\cite{Li:2021dmf,Gautason:2020tmk,Dong:2020uxp,Alishahiha:2020qza,Ling:2020laa,Matsuo:2020ypv,He:2021mst,Miao:2022mdx,Li:2023fly,Chang:2023gkt,Aguilar-Gutierrez:2023ymx,Jain:2023xta,Franken:2023ugu}.}.

\begin{figure}[H]
    \centering
    \includegraphics{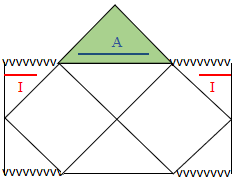}
\caption{Island region $I$ for the entropy of region $A$ in the Minkowski ‘hat’ gluing on the future boundary of inflating region(shaded in light green).
}
    \label{fig.1}
\end{figure}

In the model of Jackiw-Teitelboim (JT) gravity
\cite{Jackiw:1984je,Teitelboim:1983ux} with a positive cosmological
constant (hereafter dS$_2$ JT)
\cite{Maldacena:2019cbz,Cotler:2019nbi}:
\begin{equation}\label{eq1.2}
S=\frac{\phi_{0}}{4\pi}\int_{\cal M} \sqrt{-g}R
    +\frac{1}{4\pi} \int_{\cal M}\sqrt{-g}\phi(R-2)+S_{\partial {\cal M}},
\end{equation}
%where
%$\Sigma$ is the bulk region, $\partial\Sigma$ is the
%boundary,
%on the boundary and $K$
%is the extrinsic curvature.
%where ${\partial {\cal M}}$ is the boundary of $\cal M$,
both the crunching region and the de-Sitter (dS) region, i.e.
inflating patch, coexist. In the case where a Minkowski ‘hat’
is glued on the inflating patch, there could be an island in the
black hole patch encoded a region in the Minkowski spacetime
\cite{Hartman:2020khs,Chen:2020tes}, see Fig.\ref{fig.1}. This
$\mathrm{dS_{2}}$ spacetime can be extended to the model with many
alternating inflating and crunching regions, and the corresponding
model is a multiverse model,
e.g.\cite{Aguilar-Gutierrez:2021bns,Levine:2022wos}.

\begin{figure}[H]
    \centering
    \includegraphics[width=16cm]{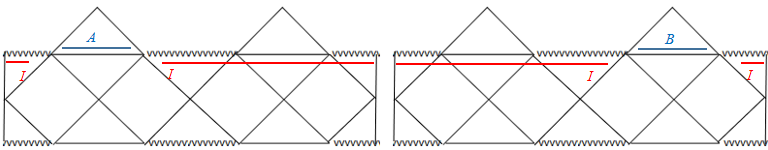}
\caption{A multiverse with two copies of spacetime. It seems that
the island regions of two neighbouring hats, $A$ and $B$, are
overlapped.} \label{fig.2}
\end{figure}

However, as has been mentioned in \cite{Levine:2022wos}, such a
multiverse model leads to a paradox as follows. In such a model
where two Minkowski bubbles are glued on two different inflating
patches with a crunching region in between, we consider two
observers - Alice and Bob - locating on two bubble regions,
respectively. Then we consider the island encoded in each region,
see Fig.\ref{fig.2}, and find it explicitly overlapped. Thus both
Alice and Bob can encode information in the overlapping region,
indicating a violation of \textsf{no-cloning theorem of quantum
mechanics}. In Ref.\cite{Levine:2022wos}, one resolution to this
paradox has been proposed\footnote{They modify the JT action to
insert a canonical singularity at a proper point, and find a
result that the dominant saddle of the gravitational path integral
is Alice and Bob's region locates on their own spacetime, which
suggests that there is no overlapping island region. }, see
\cite{Aguilar-Gutierrez:2021bns,Baek:2022ozg} for other relevant
perspectives.

In this paper, we would like to present a different resolution to
this paradox. In our scenario, the island regions which are
encoded in the region of observers in the Minkowski hat should not
include the rest of the universe, as showed in Fig.\ref{fig.2}.
Instead, the island region of a hat can only extend halfway into
the ``overlapping" region, while the other part of the
``overlapping" region entangles with the neighbouring hat. As a
result the island regions of a hat are bounded by ``division
points" rather than extending to the rest of the entire spacetime.
Therefore, there is no overlapping, Alice and Bob can only encode
information in the part of island that separately entangles with
their own patch.

%The following question is: how can we derive the new points that
%bound the islands?

We will derive ``division points" that bound the islands using
Petz $\mathrm{R\acute{e}nyi}$ mutual information (PRMI). In
section \ref{2}, we focus on the multiverse model in
$\mathrm{dS_{2}}$ JT spacetime with Minkowski bubbles glued on the
future boundaries of inflating regions, in particular the QES of
islands and the generalized entropy without the backreaction
effect. In section \ref{5}, we present our resolution to the
paradox. In section \ref{7}, we discuss the implication of our
bounded island for the traversable wormhole model in
$\mathrm{dS_{2} }$ JT spacetime. In section \ref{6}, we
summarize our results and discuss the outlook of future
researches.

%with the method of mutual information:
%\begin{equation}\label{eq1.3}
%    I_{vN}(A;B)=S_{vN}(A)+S_{vN}(B)-S_{vN}(AB).
%\end{equation}
%This is the formula of von-Neumann mutual information between the
%two subsystem $A$ and $B$. However, when it comes to
%$\mathrm{R\acute{e}nyi}$ mutual information, we need to write it
%in a more generalized form:
%\begin{equation}\label{eq1.4}
%    I_{\alpha}(A;B)=D_{\alpha}(\rho_{AB}
%    \Vert
%    \rho_{A}
%    \otimes
%    \rho_{B}),
%\end{equation}
%where $\rho $ is the density matrix,
%$D_{\alpha}(\rho \Vert \sigma) $
% is called Petz $\mathrm{R\acute{e}nyi}$ relative entropy (PRRE)
% \cite{1986Quasi}
% and $I_{\alpha}(A;B)$ is called Petz Renyi mutual information (PRMI). We will review their properties and calculations in appendix
% \ref{4}
% and they will help us in our calculation of the ‘division points’ in section
% \ref{5}. After that, we would like to take the traversable wormhole \cite{Maldacena_2013,Freivogel_2019,Gao_2017} in $\mathrm{dS_2}$ JT model as an example and discuss the new understandings these new boundary points may provide us about the relation of the entanglement island and traversable wormhole in section \ref{7}.

\section{Islands in dS$_{2}$ JT multiverse model}\label{2}

In this section, following
Refs.\cite{Aguilar-Gutierrez:2021bns,Levine:2022wos}, we construct a
multiverse model of $\mathrm{dS_{2} }$ JT gravity with Minkowski
‘hats’ glued on it.

The global coordinate of $\mathrm{dS_{2} }$ spacetime is:
\begin{equation}\label{eq2.1}
ds^2=\frac{-d\sigma^2+d\varphi^2}{\cos^2{\sigma}},
\quad
\sigma\in(-\frac{\pi}{2},\frac{\pi}{2}),
\quad
\varphi\in(-\pi,\pi).
\end{equation}
Then, varying (\ref{eq1.2}), we get the equation of motion:
\begin{equation}\label{eq2.2}
(g_{\mu\nu}\nabla^2-\nabla_{\mu}\nabla_{\nu}+g_{\mu\nu})\phi=2\pi
\langle T_{\mu\nu} \rangle,
\end{equation}
where $\langle T_{\mu\nu} \rangle$ is the expectation value of the
covariant stress-energy tensor of the CFT. The vacuum solution
($\langle T_{\mu\nu} \rangle=0$) in Nariai limit is
\cite{Balasubramanian:2020xqf}:
\begin{equation}\label{eq2.3}
\phi=\phi_{r}\frac{\cos\varphi}{\cos\sigma}.
\end{equation}
This solution is periodic in spacelike direction $\varphi$, so it
is natural to consider the extension in this direction. Rescaling
$\sigma=n\tilde{\sigma}$, and $\varphi=n\tilde{\varphi}$ with the
spatial coordinate $\tilde{\varphi}\in(-\pi,\pi)$, we perform a
coordinate change:
\begin{equation}\label{eq2.4}
    z=e^{-i(\tilde{\sigma}+\tilde{\varphi})},
    \quad
    \bar{z}=e^{-i(\tilde{\sigma}-\tilde{\varphi})}.
\end{equation}
As a result, the metric (\ref{eq2.1}) is:
\begin{equation}\label{eq2.5}
    ds^2=\frac{-dzd\bar{z}}{\mathrm{\Omega^2}},
    \quad
    \mathrm{\Omega}=\frac{1}{2n}(z\bar{z})^{(\frac{1-n}{2})}(1+(z\bar{z}^n)).
\end{equation}
According to Ref.\cite{Calabrese:2009qy}, we have the entanglement
entropy of CFT:
\begin{equation}\label{eq2.6}
    S_{\mathrm{CFT}}=\frac{c}{6}\log(\frac{(z_2-z_1)(\bar{z_2}-\bar{z_1})}{\mathrm{\Omega}\mathrm{\bar\Omega}})
    =\frac{c}{6}\log(\frac{2n^2(\cos(\frac{\sigma_2-\sigma_1}{n})-\cos(\frac{\varphi_2-\varphi_1}{n}))}{\epsilon^2\cos\sigma_2 \cos\sigma_1}),
\end{equation}
where $\epsilon$ is the ultraviolet cutoff and $c$ is the central
charge. It is noteworthy that the Weyl anomaly and Casimir energy
of CFT will contribute $\langle T_{\mu\nu} \rangle$, which is
\cite{Levine:2022wos}:
\begin{equation}\label{eq2.7} \langle
T_{\pm\pm}\rangle=\frac{c}{24\pi}(1-\frac{1}{n^2}),
\end{equation}
when we set $x_{\pm}=\sigma\pm\varphi$ and
$e^{-2\omega}=\cos^2(\sigma)$. It is explicit that it will be
cancelled when $n=1$. However, in multiverse model it will lead to
a correction to the dilaton solution (\ref{eq2.3}):
\begin{equation}\label{eq2.8}
    \phi=\phi_{r}\frac{\cos\varphi}{\cos\sigma}
    -\frac{c}{12\pi}(1-\frac{1}{n^2})(1+\sigma\tan\sigma).
\end{equation}
In this paper, for the simplicity of calculation we set
$\phi_{r}\gg c$, so for the rest of the paper we will neglect the
backreaction effect and (\ref{eq2.3}) can be still regarded as the
dilaton solution.

Then we glue a Minkowski hat on each inflation region (we refer
the Minkowski region as hat). We assume that this welding is near
the future boundary of the inflating region,
$\sigma(\varphi)=\frac{\pi}{2}-\delta\sigma$, then the boundary
condition in the global coordinate is:
%\begin{equation}\label{eq2.9}
    $\phi_{r}\frac{\cos\varphi}{\delta\sigma}=\frac{\tilde{\phi}}{\epsilon}$,
%\end{equation}
%The RHS of the equation is the dilaton of Minkowski hat. Then we
%get the $\delta\sigma$:
which suggests %\begin{equation}\label{eq2.10}
    $\delta\sigma=\epsilon\frac{\phi_{r}}{\tilde{\phi_{r}}}\cos\varphi$.
%\end{equation}
In the light of Milne wedge that cover a portion of inflating
region and the hat: %\begin{equation}\label{eq2.11}
$\tanh\chi=\frac{\sin\varphi}{\sin\sigma}$ and
    $\tanh\eta=\frac{\cos\sigma}{\cos\varphi}$,
%\end{equation}
we can fix the metric of the hat:
\begin{equation}\label{eq2.12}
    ds^2=\frac{-d\eta^2+d\chi^2}{\eta_0^2},
\end{equation}
where
\begin{equation}\label{eq2.13}
    \tanh\eta_0\approx\eta_0=
    \frac{\cos(\frac{\pi}{2}-\delta\sigma(\varphi))}{\cos\varphi}\approx
    \epsilon\frac{\phi_{r}}{\tilde{\phi_{r}}}.
\end{equation}

\begin{figure}[H]
    \centering
    \includegraphics[width=16cm]{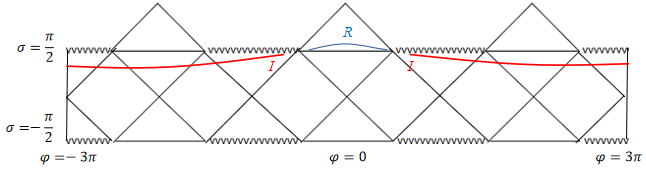}
    \caption{Islands of region $R$ in $n=3$ multiverse spacetime.}
    \label{fig.3}
\end{figure}

Rewrite
(\ref{eq2.12})
in the global coordinate:
\begin{equation}\label{eq2.14}
    ds^2=\frac{-d\sigma^2+d\varphi^2}{\eta_0^2 (\cos^2(\varphi)-\cos^2(\sigma))}
    =\frac{dzd\bar{z}}{\mathrm{\tilde{\Omega}}\mathrm{\bar{\tilde{\Omega}}}},
    \tilde{\Omega}=\frac{\eta_0 e^{i\tilde{\sigma}}}{n} \sqrt{\cos^2(\varphi)-\cos^2(\sigma)}.
\end{equation}
By replacing $\Omega$ with $\tilde{\Omega}$ in (\ref{eq2.6}),
together with the dilaton solution, we get the generalized entropy
of a hat with island regions in the crunching region:
\begin{equation}\label{eq2.15}
    S_{\mathrm{gen}}
 =\frac{c}{3}\log(\frac{2n^2(\cos(\frac{\sigma_2-\sigma_1}{n})-\cos(\frac{\varphi_2-\varphi_1}{n}))}{\epsilon^2\cos\sigma_2
 \sqrt{\cos^2(\varphi)-\cos^2(\sigma)}})+2\phi_{r}\frac{\cos\varphi}{\cos\sigma}+2\phi_{0},
\end{equation}
where the constant $\eta_0$ has been absorbed into $\epsilon$,
since it will not affect the position of the endpoints of islands.
It is speculated that the endpoints of $R$ and island regions are
close to the corners of their patches (see Fig. \ref{fig.3}):
%\begin{equation}\label{eq2.16}
$\sigma_R=\frac{\pi}{2}+\delta\sigma_R$,
$\varphi_R=\frac{\pi}{2}-\delta\varphi_R$,
$\sigma_I=\frac{\pi}{2}-\delta\sigma_I$ and
$\varphi_I=\frac{\pi}{2}+\delta\varphi_I$,
%\end{equation}
thus we have
\begin{equation}\label{eq2.17}
    S_{\mathrm{gen}}
    =\frac{c}{3}\log(\frac{(\delta\varphi_I + \delta\varphi_R)^2-(\delta\sigma_I-\delta\sigma_R)^2}{\epsilon^2 \delta\sigma_I \sqrt{(\delta\varphi_R)^2-(\delta\sigma_R)^2}})-2\phi_{r}\frac{\delta\varphi_I}{\delta\sigma_I}+2\phi_{0}.
\end{equation}
%We assume $\delta\sigma_I \gg \delta\sigma_R$,
The island solution suggests:
\begin{equation}\label{eq2.18}
\delta\sigma_I=\frac{6\phi_{r}}{c}\delta\varphi_R, \quad
\delta\varphi_I=\sqrt{1+\frac{36(\phi_{r})^2}{c^2}}
\delta\varphi_R\thickapprox \frac{6\phi_{r}}{c}\delta\varphi_R,
\end{equation}
where we have set $\phi_{r}\gg c$ and assumed $\delta\sigma_I \gg
\delta\sigma_R$. As a result, when region $R$ nearly covers the
entire hat, it seems that its island regions cover the rest of
entire spacetime, see Fig.\ref{fig.3}.

%As we have set $\phi_{r}\gg c$, the solutions (\ref{eq2.18}) can
%be approximated to:
%\begin{equation}\label{eq2.19}
%    \delta\sigma_I=\delta\varphi_I=\frac{6\phi_{r}}{c}\delta\varphi_R.
%\end{equation}

\section{A resolution to the violation of no-cloning theorem}\label{5}

In the case of Fig.\ref{fig.3}, we have a paradox: both Alice and
Bob can encode information in the overlapping region, indicating a
violation of \textsf{no-cloning theorem}. The origin of this
paradox is that the entanglement wedge of a hat can extend to the
rest of entire spacetime, which results the inevitable overlap of
the island regions of different hats.
%To avoid the paradox, this premise should be wrong.
Here, we will show that instead of extend to the rest of entire
spacetime the other end of the island might be bounded by what we
called ``division point".

\begin{figure}[H]
    \centering
    \includegraphics[width=16cm]{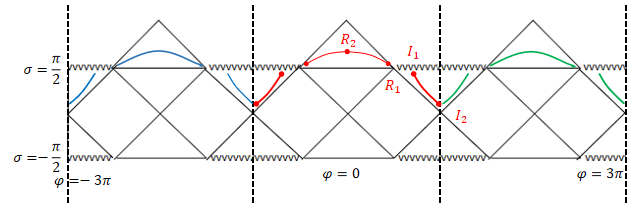}
\caption{For one hat in the multiverse model, its island regions
are bounded by the dashed lines, or we could say, the supposed
‘overlapping’ regions are divided by the dashed lines. $I_2$
can be regarded as one of the division points of the islands,
whose location is what we want to derive.
    }
    \label{fig.5}
\end{figure}

In the following, we will derive the ``division point" of the
islands. As illustrated in Fig.\ref{fig.5}, we consider the red
region centered at $\phi=0$. The region $R$ can only entangle with
the islands between the dashed lines, thus the states between the
dashed lines can be regarded as pure, i.e.the region $(I_1,I_2)
\cup (R_1,R_2)$ purifies the region $(R_1,I_1)$.

Here, we would use the PRMI $I_{\alpha}(A;B) $, which we will
review in Appendix-\ref{4}. In the limit $\alpha \to 1$, we have
\begin{equation}\label{eq5.1}
    I_{\alpha=1}(A_{(I_1,I_2)};B_{(R_1,R_2)})
    \sim
    S_{\mathrm{CFT}}(R_1,I_1),
\end{equation}
where $I_{\alpha=1}$ is difficult to calculate\footnote{It
contains a four-point function:
$\mathrm{Tr}(\rho_{AB}^{\alpha}(\rho_{A} \otimes \rho_{B})^{m})
=\langle\sigma_{g_A}(A_1)\sigma_{g_A
^{-1}}(A_2)\sigma_{g_B}(B_1)\sigma_{g_B ^{-1}}(B_2)
\rangle_{\mathbb{C}}$, which we do not have an analytic
expression.}. However, considering $d(R_1,I_1)$ is relatively
small, we could use the adjacent approximation, and slightly
modify (\ref{eq4.16}) as\footnote{The accuracy of (\ref{eq5.3})
will be discussed in Appendix-\ref{8}.}\cite{Kudler-Flam:2023kph}:
\begin{equation}\label{eq5.3}
    I_{\alpha=1}(A_{(I_1,I_2)};B_{(R_1,R_2)})=
    \frac{1}{3}\log\frac{l_A l_B}{d(l_A + l_B)},
\end{equation}
where $l_A=d(I_1,I_2)$, $l_B=d(R_1,R_2)$, and $d=d(I_1,R_1)$. We
assume the UV cut-off cancels and neglect the constant correction
factors, Eq.(\ref{eq5.1}) is approximately:
\begin{equation}\label{eq5.4}
    \frac{1}{l_A}+\frac{1}{l_B} \sim \frac{1}{d^2}_{.}
\end{equation}

In the light of (\ref{eq2.18}), the coordinates of $R_1$, $I_1$
are:
\begin{equation}\label{eq5.5}
    R_1 :(\frac{\pi}{2}+\delta\sigma_R , \frac{\pi}{2}-\delta\varphi_R),
    \quad
    I_1 : (\frac{\pi}{2}-\frac{6\phi_{r}}{c}\delta\varphi_R,\frac{\pi}{2}+\frac{6\phi_{r}}{c}\delta\varphi_R).
\end{equation}
We assume $R_2:(\frac{\pi}{2}+\delta\sigma_R,0)$ is near the
future boundary of inflating region and set $\delta\varphi_R =
N\delta\sigma_R $. According to Fig.\ref{fig.5}, we can easily
find $N>1$. Then we will get:
\begin{equation}
    \begin{aligned}
        \label{eq5.6}
    &d^2=\frac{(z_{I_1}-z_{R_1})(\bar{z}_{I_1}-\bar{z}_{R_1})}{\Omega_{{I_1}}\tilde{\Omega}_{{R_1}}}=\frac{2}{\sqrt{ 1-N^{-2}}}, \\
    &l_B ^2=\frac{(z_{R_2}-z_{R_1})(\bar{z}_{R_2}-\bar{z}_{R_1})}{\tilde{\Omega}_{{R_1}}\tilde{\Omega}_{{R_2}}}=
    \frac{2n^2(1-\cos(\frac{\frac{\pi}{2}-\delta\varphi_R}{n}))}
    {\delta\varphi_R \sqrt{ 1-N^{-2}}},
    \end{aligned}
\end{equation}
Thus we have
\begin{equation}\label{eq5.7}
    l_A = d^2,
\end{equation}
since $l_B$ is very large for $n\gg 1$. Considering that there
exists axial symmetries at $\phi=n\pi$, we could set
$I_2:(\sigma_2,\pi)$, and substitute it to (\ref{eq5.7}):
\begin{equation}\label{eq5.8}
    \frac{2n^2((\cos
        \frac{\sigma_2-
            (\frac{\pi}{2}-
                \frac{6\phi_{r}}{c}\delta\varphi_R)}{n})
        -\cos(
        \frac{
            \frac{\pi}{2}-
                \frac{6\phi_{r}}{c}\delta\varphi_R}{n}))}
    {\frac{6\phi_{r}}{c}
        \cos\sigma_2
        \delta\varphi_R }
    =
    \frac{4}{1-N^{-2}}.
\end{equation}
Though it seems rather complicated to solve (\ref{eq5.8})
directly, when $n \gg 1$ we could approximately have
\begin{equation}\label{eq5.9}
    \frac{(\frac{\pi}{2}-
        \frac{6\phi_{r}}{c}\delta\varphi_R)^2
        -
        (\sigma_2-
            (\frac{\pi}{2}-
                \frac{6\phi_{r}}{c}\delta\varphi_R))^2}
    {\frac{6\phi_{r}}{c}
        \cos\sigma_2
        \delta\varphi_R}
    =\frac{4}{1-N^{-2}}.
\end{equation}
As $\delta\varphi_R$ is very small, $\sigma \in
(-\frac{\pi}{2},\frac{\pi}{2})$, it is explicitly that $\sigma_2$
should be very close to 0, which makes $\cos\sigma_2 \sim 1$. Then
solving (\ref{eq5.9}), we get:
\begin{equation}\label{eq5.10}
    \sigma_2=\frac{24\phi_{r}}
    {\pi c ( 1-N^{-2})}\delta\varphi_R  ,
\end{equation}
which is consistent with our assumption that $\sigma_2$ should be
very small. Thus we seek out such a division point $I_2$ of
island, actually very close to the event horizon.

Therefore, the island region encoded in a hat in the dS$_{2}$
multiverse model are bounded instead of extending to the rest of
entire spacetime. In corresponding case, the paradox can be
avoided since the observer in one hat could only encoded the
information in the bounded island regions in the multiverse model
(Fig.\ref{fig.5}), The position of the boundaries can be fixed
with the help of mutual information calculations and symmetry. In
the ``bird eye" view the existence of such boundaries does not
affect the endpoints of the islands. Instead, they seems to affect
the ``shape" of the islands.

It should be commented that if we choose the ``bird eye" view of
the multiverse in corresponding model, we will still find the
island regions continuous, however, for an observer, such as
Alice, at one hat, when she attempted to decode the island region,
she has no ideas for the regions beyond the division points, and
what she can only know is the island regions bounded by the
division points in the neighbouring crunching regions,

\section{Connection with traversable wormhole}\label{7}

It is natural to consider the implications of the division points
that bound the islands. In this section we argue that it might
bring us a new perspective of understanding the relationship
between the entanglement islands and traversable wormhole.

Traversable wormhole plays an important part in ER=EPR conjecture
\cite{Maldacena:2013xja,Susskind:2017nto,Verlinde:2020upt,Bennett:1992tv},
making it possible for information transferring through the
horizon. It also serves as a resolution to the cloning paradox in
Hayden-Preskill' work \cite{Hayden:2007cs,Maldacena:2017axo}. To
construct a traversable wormhole in a crunching spacetime, we need
to impose a pulse with negative energy
\cite{Freivogel:2019lej,Gao:2016bin}.

In dS$_2$ JT spacetime with the Kruskal coordinate \footnote{The
shift from the Kruskal coordinate to the global coordinate are:
$\tan \sigma =\frac{U+V}{1-UV}$ and $\tan \varphi
=\frac{V-U}{1+UV}$.}
\begin{equation}\label{eq7.2}
    ds^2=-\frac{4}{(1-UV)^2} dUdV,
\end{equation}
a negative pulse
\cite{Aguilar-Gutierrez:2023ymx}\footnote{We assumed
$\vert \alpha \vert$ is small so that the backreaction effect to
the islands can be neglected. }:
    \begin{equation}\label{eq7.1}
    \langle T_{VV}  \rangle
    =\alpha \frac{\delta(V-V_{S})}{V_{S}},
    \quad with \quad
    \alpha < 0,
\end{equation}
will result in a shift:
\begin{equation}\label{eq7.3}
    \Delta U =-\alpha\frac{\pi }
    {V_S \phi_{r}}+\mathcal{O}(\alpha ^ 3)
    %=-\frac{4\pi \alpha \epsilon}
    %{c V_S}+\mathcal{O}(\alpha ^ 3),
\end{equation}
The computation of (\ref{eq7.3}) will be presented in
Appendix-\ref{9}. The coordinate of null ray move through
different patches will be changed by
%\begin{equation}\label{eq7.5}
    $U_i \to - \frac{1}{U_i}$ and
    %\quad
    $V_i \to - \frac{1}{V_i}$,
%\end{equation}
as the motion from the crunching region to the inflating region
result in a shift of $\phi \to \phi + \pi$. Thus for the null ray
sent from $U=U_2$, it will enter the inflating region along the
path:
\begin{equation}\label{eq7.6}
    U_{\alpha}= - \frac{1}{U_2 - \Delta U}.
\end{equation}

\begin{figure}[H]
    \centering
    \includegraphics[width=16cm]{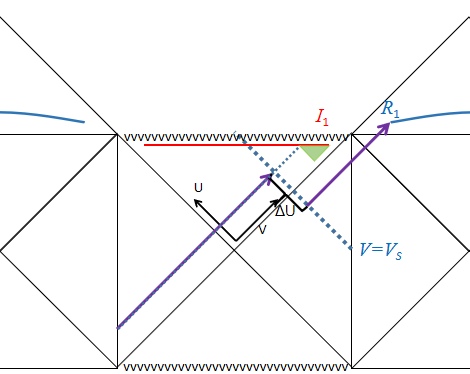}
\caption{Traversable wormhole in $\mathrm{dS_{2} }$ JT Nariai
spacetime with Minkowski hat. In convention we choose to fix the
Penrose diagram and makes the null ray (purple line) shift
$-\Delta U$. If the island is close to the singularity everywhere,
only a few of signals whose geodesic lines intersect with the
casual diamond of the island (shaded in light green) can reach the
observer in region $R$.
    }
    \label{fig.7}
\end{figure}

\begin{figure}[H]
    \centering
    \includegraphics[width=16cm]{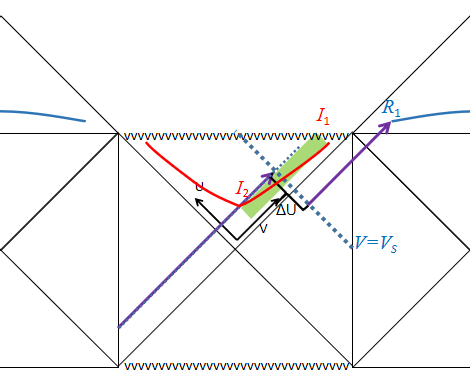}
\caption{The division point $I_2$ change the shape of the island.
Considering the pulse is small, $I_2$ should still close to the
horizon. When the null ray emitted at the north pole (purple line)
intersect with $I_2$ and can reach the endpoint of $R$, all
signals that intercept the causal diamond of right half island
(shaded in light green) can reach $R$. In this case the
traversable wormhole cover the casual diamond of the part of the
island that entangles with the half $R$.
    }
    \label{fig.8}
\end{figure}

The light ray could intersect with region $R$ for $U_{\alpha} <
U_R$. In the case of Fig.\ref{fig.7}, the traversable wormhole can
only cover a small portion of the causal diamond of the right half
of the island entangled with region $R$, as $\vert \alpha \vert$
is small, however, in the case that the division points exist (see
Fig.\ref{fig.8}), a large portion of the causal diamond of the
right part of the island entangled with the region $R$ can be
covered by the traversable wormhole.

In the following, we will present an argument into this insight.
In a case that all signals that intercept with the causal diamond
of the right half of the island can reach $R$, the coordination of
$I_2$ must obey:
\begin{equation}\label{eq7.7}
    U_R = U_{\alpha}= - \frac{1}{U_2 - \Delta U}
    =- \frac{1}{U_2 + \frac{4 \pi \alpha\epsilon}{c V_s}}.
\end{equation}
where $\epsilon=\frac{c}{4\phi_{r}}$. Thus we have
\begin{equation}\label{eq7.8}
    I_2:\left(\frac{12 \phi_{r}}{(1-N^{-2})c} \delta \varphi_R,
    -\pi\right),
\end{equation}
where $n=1$ is set. We assume the pulse is imposed at
$\sigma_S=0$, so that we can fix $V_S=1 $. Thus the positions of
$I_2$, $R_1$ in the Kruskal coordinate are:
\begin{equation}\label{eq7.10}
    \begin{aligned}
    & U_2 = V_2 = \frac{6 \phi_{r}}{
        ( 1-N^{-2})  c} \delta \varphi_R
    = \frac{3 \delta \varphi_{R}}{2 (1-N^{-2} )\epsilon}, \\
    & U_R \sim \frac{2}{\delta \varphi_R}
    \qquad
    V_R \sim \frac{\delta \varphi_R}{2}.
    \end{aligned}
\end{equation}
Thus substituting (\ref{eq7.10}) to (\ref{eq7.7}), we have
\begin{equation}\label{eq7.11}
    \alpha = -\frac{c}{8 \pi \epsilon}
    \left(\frac{3}{(1-N^{-2} )\epsilon}-1\right)\delta \varphi_R.
\end{equation}
As expected, $|\alpha|$ is very small. However, if the amplitude
of pulse is bigger, we cannot ignore the backreaction effects, and
the location of island will be shifted, see (\ref{eq9.7}) in
Appendix-\ref{9}. In that case, the shift of the island in $U$
direction cancels the shift of the null ray caused by the pulse,
since the position of shifted division point $I_2$ (for $V_S=1$)
is:
\begin{equation}\label{eq7.13}
	\begin{aligned}
    U_2 = V_2
    =\frac{3 \delta \varphi_{R}}{2 (1-N^{-2}
    )\epsilon}-\alpha\frac{\pi}{\phi_r}
    & =\frac{3 \delta \varphi_{R}}{2 (1-N^{-2}
    )\epsilon} +\Delta
    U>\Delta
    U,
    \\
    U_1=\frac{V_r \epsilon}{3}
    &+\Delta U > \Delta U
	\end{aligned}
\end{equation}
it is impossible for the signals that pass through the causal
diamond of island to travel through the horizons. Thus when a
pulse get too large, it will favor to close the traversable
wormhole in dS$_2$ spacetime, see also Ref.\cite{Maldacena:2017axo}
for AdS spacetime.

%ER=EPR conjecture indicates a possibility of information
%transformation between entangled systems through the wormhole
%\cite{Susskind_2018}, and if we allow interaction between the
%entangled regions, which is realized by introduced a pulse in this
%situation, we will be able to make the wormhole traversable
%\cite{Gao_2017}.

In our case the traversable wormhole has causal contact with
portion of the entanglement island, indicating a possible
potential connection between them. The division point we derived
in section.\ref{5} can make the traversable wormhole cover most of
the casual diamond of the half part of the island that entangles
with the observer, which might strengthen the correlation between
the traversable wormhole and ER=EPR conjecture.

%This may lead to another interesting phenomenon that part of the
%island may be seen directly from the entangled observer through
%the traversable wormhole which only requires a relatively small
%pulse.

\section{Discussions}\label{6}

One of the major differences between the cosmology and black holes
is that cosmological event horizons are observer-dependent. Two
observers outside a black hole will agree one same event horizon,
and they cannot encode the same interior region at once. However,
two observers in different patch of the universe may have
different event horizons. They can encode the regions beyond their
horizons, but these regions seems inevitably overlap, which
violates the \textsf{no-cloning theorem}.

In this paper, our resolution to this paradox is that the regions
beyond the horizons are bounded so that they cannot overlap. In
implementing detail, we focus on a multiverse model in
$\mathrm{dS_{2} }$ JT gravity, and set a proper condition so that
we can neglect the backreaction effect to simplify the
calculation. In corresponding model, with the help of the PRMI and
symmetry we find that while one end of the island region of one
hat is bounded by the QES, the other end of the island is bounded
by what we called ``division point" instead of extend to the rest
of entire spacetime. Bounded by the division points, the island
regions of different hats do not overlap any longer, thus the
paradox can be avoided, see Fig.\ref{fig.5}.

It is worth mentioning that although the division points seem to
divide the island regions apart, they do not actually physically
divide the islands. In the ``bird eye" view the island regions are
still continuous while the regions for an observer on a hat can
decode end at the division points. The difference of their
understanding for islands originates from their different view,
one is at a Minkowski hat, the other is in the ``bird eye" view,
though the latter perspective of view might not actually exist in
reality.

The existence of the division points affect the ``shape" of the
islands, which might be an interesting bonus. It is well known
that the endpoints of the island are usually fixed by the QES, and
we seldom care about the shape of the island region itself. In
this paper we discuss the role of division points might play in
the traversable wormhole model in $\mathrm{dS_2}$ spacetime, which
shows that the division points of the island might actually affect
the shape of the causal diamond of half of the island that
entangles with the outside observer, just as Fig.\ref{fig.8}.

%the traversable wormhole can fully cover the width of causal
%diamond of the half of the island even if the pulse is relatively
%small! This may provide a new perspective of interpreting the
%ER=EPR conjecture between the entanglement island and the related
%radiation outside the horizon.
%We expect that there might be more interesting properties of these
%bounded islands that remain to be discovered.

Here, we focus on a $\mathrm{dS_{2} }$ JT multiverse model, it
might be interesting to consider whether our results can be
extended into higher dimensions or more realistic spacetime,
e.g.\cite{Farhi:1989yr}, or \cite{Li:2009me} and recent e.g.
\cite{Huang:2023chx,Huang:2023mwy} for inflating spacetime with
supermassive primordial black holes. To simplify the calculation,
we have set $\phi_r \gg c$ to neglect the backreaction effect, we
wonder with the backreaction if there would be some significant
differences. And we make an adjacent approximation to PRMI, but in
fact it is a two disjoint intervals calculation, which depends on
the full operator content of the theory, and its expansion is
complicated and depends on the specific models
\cite{Calabrese:2010he,Cardy:2013nua}. It is worthwhile to explore
relevant issues.

%Furthermore, PRRE itself can be generalized into $\alpha-z$
%$\mathrm{R\acute{e}nyi}$ relative entropy \cite{2015α}, and the
%replica manifold of related $\alpha-z$ $\mathrm{R\acute{e}nyi}$
%mutual information is more complicated than that of PRMI
%\cite{kudlerflam2023renyi}. We expect more researches on
%$\alpha-z$ $\mathrm{R\acute{e}nyi}$ relative entropy might provide
%new understandings of black hole and cosmological information
%issues.

\textbf{Acknowledgment} This work is supported by NSFC,
No.12075246, National Key Research and Development Program of
China, No. 2021YFC2203004, and the Fundamental Research Funds for
the Central Universities.

\appendix

\section{On PRRE and PRMI}\label{4}

In this Appendix, we briefly review some properties and
calculations of Petz $\mathrm{R\acute{e}nyi}$ relative entropy
(PRRE) and Petz $\mathrm{R\acute{e}nyi}$ mutual information
(PRMI).

The well known von-Neumann entropy is:
\begin{equation}\label{eq4.1}
    S_{vN}=-\mathrm{Tr}(\rho\log\rho),
\end{equation}
where $\rho$ is the density matrix. To measure the correlation
between two subsystems $A$ and $B$, we would take the mutual
information:
\begin{equation}\label{eq4.2}
    I_{vN}(A;B)=S_{vN}(A)+S_{vN}(B)-S_{vN}(AB).
\end{equation}

However, with the $\mathrm{R\acute{e}nyi}$ entropy:
\begin{equation}\label{eq4.4}
    S_n(\rho)=\frac{1}{1-n}\log(\mathrm{Tr}\rho^n),
\end{equation}
the $\mathrm{R\acute{e}nyi}$ mutual information (RMI) need to be
written as \cite{Kudler-Flam:2023kph}:
\begin{equation}\label{eq4.7}
    I_{\alpha}(A;B)=D_{\alpha}(\rho_{AB}\Vert \rho_{A} \otimes \rho_{B}),
    \quad with \quad
    D_{\alpha}(\rho \Vert \sigma)
    =\frac{1}{\alpha-1}\log[\mathrm{Tr}(\rho^{\alpha}\sigma^{1-\alpha})],
\end{equation}
which is called PRMI, while $D_{\alpha}$ is called Petz
$\mathrm{R\acute{e}nyi}$ relative entropy (PRRE)
\cite{Petz:1986naj,1973Convex,1977Relative}. $D_{\alpha}$ will return
to von-Neumann entropy at the limit of $\alpha \to 1$
\cite{Kudler-Flam:2021alo}.

In analogy to the situation of $\mathrm{R\acute{e}nyi}$ entropy,
we use replica trick to calculate PRMI
\cite{Kudler-Flam:2021alo,Kudler-Flam:2023kph}:
\begin{equation}\label{eq4.8}
    I_{\alpha}(A;B)
    =\frac{1}{\alpha-1}\lim_{m \to 1-\alpha}\log[\mathrm{Tr}(\rho_{AB}^{\alpha}(\rho_{A} \otimes \rho_{B})^{m})].
\end{equation}

\begin{figure}[H]
    \centering
    \includegraphics[width=16cm]{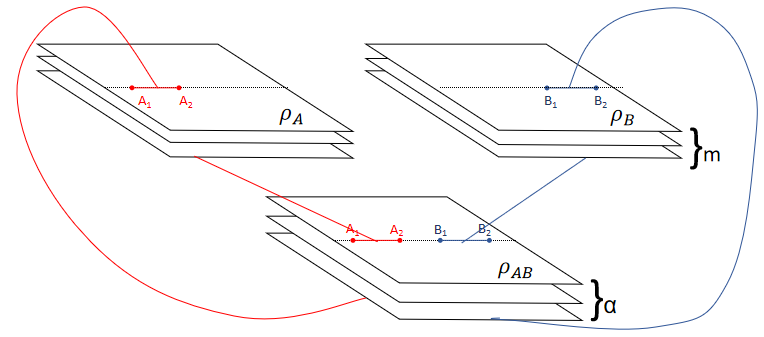}
    \caption{The replica manifold of
$\mathrm{Tr}\log((\rho_{AB})^{\alpha}(\rho_{A} \otimes
\rho_{B})^m) $, the lines correspond to the structure of gluing
between the replicas.
        }
    \label{fig.4}
\end{figure}

Here, similar to the method of \cite{Calabrese:2009qy}, for the
replica manifold of PRRI shown in Fig. \ref{fig.4}, we can also
derive its scaling dimensions. One of the major differences is
that we will use the cyclic permutations $g_A$, $g_B$, instead of
the cyclic permutations $(1...n)$, $(n...1)$
\cite{Kudler-Flam:2023kph}:
\begin{equation}
    \begin{aligned}
        \label{eq4.12}
g_A=(1,...,\alpha,\alpha+1,...,\alpha+m),\quad
        g_B=(1,...,\alpha,\alpha+m+1,...,\alpha+2m),
        \end{aligned}
\end{equation}
noting that the permutation $g_A ^{-1}g_B $ does not corresponds
to the identity permutation:
\begin{equation}\label{eq4.13}
    g_A ^{-1}g_B=(1,\alpha+m,...,\alpha+1,\alpha+m+1,...,\alpha+2m).
\end{equation}

Generally, the scaling dimension can be written as:
\begin{equation}\label{eq4.14}
    \Delta_g=\sum_{g \in \mathrm{cycle}(g)}
    \frac{c}{12}(\vert g_i \vert - \frac{1}{\vert g_i \vert}),
\end{equation}
then the scaling dimensions of the corresponding twists operators
can be calculated:
\begin{equation}
    \begin{aligned}
        \label{eq4.15}
        &\Delta=\Delta_{g_A}=\Delta_{g_B}=\Delta_{g_A ^{-1}}    =\Delta_{g_B ^{-1}}=\frac{c}{12}(\alpha+m-\frac{1}{\alpha+m}),\\
        &\Delta_{g_A ^{-1}g_B}=\Delta_{g_B ^{-1}g_A}=\frac{c}{12}(2m+1-\frac{1}{2m+1}).     \end{aligned}
\end{equation}

Assuming that the region $A$ has a finite length $l_A$ with
endpoints at $A_1$ and $A_2$, and the region $B$ has a finite
length $l_B$ with endpoints at $B_1$ and $B_2$, we consider a
special situation where $A$ and $B$ are adjacent, equivalently
$A_2 = B_1$. Then we can calculate the PRMI of $A$ and $B$:
\begin{equation}
    \begin{aligned}
        \label{eq4.16}
        I_{\alpha}(A;B)
        &=\frac{1}{\alpha-1}\lim_{m \to 1-\alpha}\log
        \langle\sigma_{g_A}(A_1)\sigma_{g_A ^{-1}g_B}(A_2)\sigma_{g_B ^{-1}}(B_2) \rangle_{\mathbb{C}} \\
        &=\frac{1}{\alpha-1}\lim_{m \to 1-\alpha}\log
        \frac{C_{\alpha,m}}{(l_A l_B)^{\Delta_{g_A ^{-1}g_B}}(l_A + l_B)^{2\Delta - \Delta_{g_A ^{-1}g_B}}} \\
        &=c\frac{2-\alpha}{3(3-2\alpha)}\log
        \frac{l_A l_B}{\epsilon^2 (l_A + l_B)} + \mathcal{O}(1)
        ,   \end{aligned}
\end{equation}
where $C_{\alpha,m}$ is the operator product expansion (OPE)
coefficient and leads to the $\mathcal{O}(1)$ corrections which
have no an general analytic expression. However, in some certain
situations, such as fermion systems, it can be calculated
\cite{Kudler-Flam:2023kph}.

\section{A check of accuracy on (\ref{eq5.3})}\label{8}
In this Appendix, we explore the condition that our approximation
in the calculation of PRMI (\ref{eq5.3}) in section.\ref{5} is
accurate in $n=1$ case.

\begin{figure}[H]
    \centering
    \includegraphics[width=10cm]{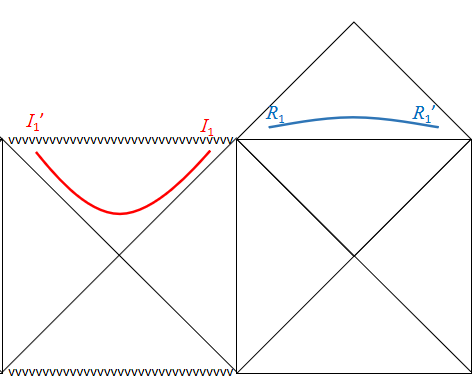}
\caption{The Penrose diagram of $\mathrm{dS_{2} }$ JT Nariai
spacetime with Minkowski hat, corresponding to the $n=1$ situation
of the multiverse model we focus.
    }
    \label{fig.6}
\end{figure}

As is shown in Fig.\ref{fig.6}, it is explicitly that $(R_1,R_1')
\cup (I_1,I_1')$ purifies the region $(I_1,R_1) \cup (I_1',R_1')$,
which leads to:
\begin{equation}\label{eq8.1}
    I_{\alpha=1}(A_{(I_1,I_1')};B_{(R_1,R_1')})
    \sim
    2S_{\mathrm{CFT}}(R_1,I_1)  ,
\end{equation}
with the coordinates of the points:
\begin{equation}\label{eq8.2}
    \begin{aligned}
        I_1(\frac{\pi}{2}-\delta \sigma_I,
        -\frac{\pi}{2}-\delta \varphi_I),
        \quad & \quad
        I_1'(\frac{\pi}{2}-\delta \sigma_I,
        -\frac{3\pi}{2}+\delta \varphi_I), \\
        R_1(\frac{\pi}{2}+\delta \sigma_R,
        -\frac{\pi}{2}+\delta \varphi_R),
        \quad & \quad
        R_1'(\frac{\pi}{2}+\delta \sigma_R,
        \frac{\pi}{2}- \delta \varphi_R),
    \end{aligned}
\end{equation}

According to our procedures in section.\ref{5}, the deviation of
our approximation will be:
\begin{equation}\label{eq8.3}
    \Delta S = \frac{1}{3}\log(d^3 (\frac{1}{l_A}+\frac{1}{l_B})),
\end{equation}
where:
\begin{equation}\label{eq8.4}
    \begin{aligned}
        d^2
        &=d^2(I_1,R_1)
        =\frac{4}{ 1-N^{-2}}, \\
                l_A^2
        &= d^2(I_1,I_1')
        =\frac{2(1-\cos(\pi-2\delta \varphi_I))}{\delta \sigma_I^2}, \\
        &=\frac{4\sin^{2}(4\delta \varphi_I)}{\delta \sigma_I^2}
        \sim 64 ,\\
        l_B^2
        &= d^2(R_1,R_1')
        =\frac{2(1-\cos(\pi-2\delta \varphi_R))}{\delta \varphi_R^2-\delta \sigma_R^2} ,\\
        &=\frac{4\sin^{2}(4\delta \varphi_R)}{\delta \varphi_R^2-\delta \sigma_R^2}
            \sim \frac{64}{1-N^{-2}}.
    \end{aligned}
 \end{equation}
Substituting to (\ref{eq8.3}), we get the deviation function:
 \begin{equation}\label{eq8.5}
    \Delta S(N)=\frac{1}{3}\log(\frac{1+( 1-N^{-2})^{\frac{1}{2}}}{( 1-N^{-2})^{\frac{3}{2}}}),
\end{equation}
which lowers monotonically with $N$ when $N>1$. Therefore, in
$n=1$ situation, our assumption will get more accurate as $N$ get
bigger. When $N\to \infty$, $\Delta S \to \mathcal{O}(1)$,
indicating that our assumption is quite accurate in this case.

\section{Derivation of traversable wormhole in $\mathbf{dS_{2} }$ JT spacetime}\label{9}

In this Appendix, we borrow the methods from
\cite{Aguilar-Gutierrez:2023ymx} to derive the shift of
the horizon and the back-reacted island.

The dilaton satisfies:
\begin{equation}\label{eq9.1}
    \phi=\phi_0 -\phi_{r} \frac{1+U_I V_I}{1-U_I V_I}-\frac{2\pi\alpha(1-V_S U_I)}{V_S (1-V_I U_I)} (V_I-V_S)\theta (V_I - V_S),
\end{equation}
where $\theta $ is the Heaviside function. To derive the shifted
horizon we need to extremalize the dilaton in $U$ and $V$
directions:
\begin{equation}\label{eq9.2}
    U^+ \sim \frac{\phi_{r}}{\pi V_S \alpha},
    \quad
    U^- \sim -\frac{\pi \alpha}{ V_S \phi_{r}}.
\end{equation}
Consider $\vert \alpha \vert$ is small, the shift of horizon
should not be very large, and it is explicitly that when $\alpha =
0$ the horizon lies at $U=0$, so the shift of horizon in $U$
direction will be:
\begin{equation}\label{eq9.3}
    \Delta U =-\frac{\pi \alpha}
{V_S \phi_{r}},
\end{equation}
which results in (\ref{eq7.3}).

To calculate the back-reacted island we need to map the Kruskal
coordinate into the vacuum coordinate $z$ and $\bar {z}$, which
can be fixed by:
\begin{equation}\label{eq9.4}
    \langle T_{VV}  \rangle
    = -\frac{c}{24 \pi} \{z,V \}.
\end{equation}
We set:
\begin{equation}\label{eq9.5}
    z=V+ \alpha f(V),
    \quad
    \bar{z}=U,
\end{equation}
in $\vert \alpha \vert \ \ll 1$, and thus the solution to
(\ref{eq9.4}) is:
\begin{equation}\label{eq9.6}
    f(V)=-\frac{12\pi}{cV_S}(V-V_S)^2 \theta(V-V_S).
\end{equation}

Then we substitute it together with the back-reacted dilaton
(\ref{eq9.1}) to the formula of generalized entropy and derive the
QES in the case $\epsilon \ll \vert \alpha \vert \ll 1$, we will
have the island affected by the backreaction at leading order of
$\epsilon$ and $\alpha$:
\begin{equation}\label{eq9.7}
    \begin{aligned}
        V_i
        &= \frac{U_r \epsilon}{3}+4\pi \alpha
        \frac{V_S \epsilon}{c}, \\
        U_i
        &= \frac{V_r \epsilon}{3}-4\pi \alpha
        \frac{\epsilon}{c V_S}.
    \end{aligned}
\end{equation}
In the case of $\alpha =0$, we have:
\begin{equation}\label{eq9.8}
    \begin{aligned}
        -\frac{1}{\delta \sigma_R}
        &=\frac{U_R+V_R}{1-U_R V_R},
        \qquad
        \frac{1}{\delta \sigma_I}
        =\frac{U_I+V_I}{1-U_I V_I}, \\
        -\frac{1}{\delta \varphi_R}
        &=\frac{V_R-U_R}{1-U_R V_R},
        \qquad
        \frac{1}{\delta \varphi_I}
        =\frac{V_I-U_I}{1-U_I V_I}.
    \end{aligned}
\end{equation}
Solving it and we will get:
\begin{equation}\label{eq9.9}
\delta \sigma_I \sim \delta \varphi_I \sim
\frac{6 \phi_{r}}{c} \delta \varphi_R,
\end{equation}
which is incoherence with the island we derive in section.\ref{2}.

\bibliography{references}

\end{document}